\documentclass[12pt]{article}

\newtheorem{pro}{Proposition}

\usepackage{times,epsfig,graphics,amssymb,latexsym,amsmath,setspace,fullpage}
\usepackage{array,threeparttable,hyperref,url}
\usepackage[usenames]{color}

\def\boxit#1{\vbox{\hrule\hbox{\vrule\kern6pt\vbox{\kern6pt#1\kern6pt}\kern6pt\vrule}\hrule}}

\def\argmax{\mathop{\rm argmax}}
\def\argmin{\mathop{\rm argmin}}

\def\sign{\mathop{\rm sign}}

\begin{document}
\pagenumbering{arabic}

\setcounter{page}{1}

\title{\Large \bf A model-free estimation for the covariate-adjusted Youden index and its associated cut-point}

\author{
 Tu Xu$^{\dag\ddag}$, Junhui Wang$^{\ddag \natural}$ and Yixin Fang$^\sharp$ \\[10pt]
     $^\dag$ Amgen Inc. \\
          \and
          $^\ddag$ Department of Mathematics, Statistics, \\
               and Computer Science \\
               University of Illinois at Chicago \\
               \and
               ~~~~~~~$^\natural$ Department of Mathematics~~~~~~~~ \\
          City University of Hong Kong \\
     \and
          ~~~~~~~~~~$^\sharp$ Division of Biostatistics~~~~~~~~~~\\
     Department of Population Health\\
     New York University}
\date{}
\maketitle

\begin{abstract}
In medical research, continuous markers are widely employed in diagnostic tests to distinguish diseased and non-diseased subjects. The accuracy of such diagnostic tests is commonly assessed using the receiver operating characteristic (ROC) curve. To summarize an ROC curve and determine its optimal cut-point, the Youden index is popularly used. In literature, estimation of the Youden index has been widely studied via various statistical modeling strategies on the conditional density. This paper proposes a new model-free estimation method, which directly estimates the covariate-adjusted  cut-point without estimating the conditional density. Consequently, covariate-adjusted Youden index can be estimated based on the estimated cut-point. The proposed method formulates the estimation problem in a large margin classification framework, which allows flexible modeling of the covariate-adjusted Youden index through kernel machines. The advantage of the proposed method is demonstrated in a variety of simulated experiments as well as a real application to Pima Indians diabetes study.

\end{abstract}


\doublespacing

\section{Introduction}

In medical research, continuous markers are widely employed in diagnostic tests to distinguish diseased and non-diseased subjects \cite{Shapiro1999}. A subject is diagnosed as diseased if the marker value is higher than a given threshold value, and otherwise non-diseased. The diagnosis accuracy of the marker is usually evaluated through sensitivity and specificity, or the probabilities of a true positive and a true negative for any given threshold value. In addition, the receiver operating characteristic (ROC) curve is defined as sensitivity versus 1-specificity over all possible threshold values for the marker \cite{Zhou2002,Pepe2003}. To summarize the overall property of an ROC curve, different summarizing indices are proposed, including the Youden index \cite{Youden1950} and the area under the ROC curve (AUC; \cite{Bamber1975}).

The Youden index, defined as the maximum vertical distance between the ROC curve and the $45^{\circ}$ line, is an indicator of how far the curve is from the uninformative test \cite{Pepe2003}. It ranges from 0 for the uninformative test to 1 for an ideal test. The Youden index has been successfully applied in many medical studies to provide an appropriate one-dimensional summary of the test accuracy and determine its associated  cut-point (e.g., \cite{Aoki1997,Castle2003}).

In literature, various statistical modeling strategies have been proposed to estimate the Youden index and its associated  cut-point. One main strategy is to use parametric models (e.g., \cite{Zou2000,Faraggi2002,Schisterman2005}), which assume that the values of the diagnostic marker for the diseased and non-diseased subjects, respectively, follow certain probability distribution. With the parametric probability distributions, explicit formulas for the Youden index and its associated cut-point can be derived. Another popular strategy uses non-parametric models (e.g., \cite{Hsieh1996,Zou1998,Knight2000}), which estimate the conditional distribution of the diagnostic marker for the diseased and non-diseased subjects via non-parametric density estimation techniques. Fluss et al. \cite{Fluss2005} conducted numerical comparisons among a number of popular estimation methods, and suggested that the kernel density estimation is generally the best performer without restricting the data distribution. With the estimated distributions of the diagnostic marker for the diseased and non-diseased subjects, the associated cut-point can be estimated as the value where the two estimated densities are identical.

Furthermore, it is important to appropriately set cut-points in subpopulations and to understand the sources of false positive and false negative results. Zhou et al. \cite{Zhou2002} and Pepe \cite{Pepe2003} discussed the covariate effect on the accuracy of diagnostic tests and the estimation of the ROC curve. Ignoring the covariate effects may lead to biased inference about the accuracy of the test for distinguishing diseased and non-diseased subjects. In data analysis, as pointed out in Pepe (\cite{Pepe2003}; page 135), ``it might be of interest to calculate both the pooled and covariate-specific ROC curves in order to ascertain the gains in accuracy that can be achieved by using covariate-specific thresholds." Although much research has been done on the covariate-adjusted ROC curve (e.g., \cite{Pepe2000,Faraggi2003,Yao2010}), little has been done on the covariate-adjusted Youden index and its associated cut-point. To the best of our knowledge, Zhou \cite{Zhou2011} studied the covariate-adjusted Youden index by using the heteroscedastic regression model \cite{Yao2010}.

In this paper, a new model-free estimation framework is proposed, which directly estimates the covariate-adjusted cut-point without estimating the conditional densities. With the estimated cut-point, the covariate-adjusted Youden index can be estimated through any one-dimensional non-parametric density estimation methods. In particular, the estimation framework formulates the estimation problem in a large margin classification framework, where the covariate-adjusted  cut-point is modeled non-parametrically in a reproducing kernel Hilbert space (RKHS;\cite{Wahba1990}). The proposed method is applied to Pima Indians diabetes study, and suggests the important effect of age in estimating the Youden index and its associated cut-point.

The rest of the paper is organized as follows. In Section 2, we briefly review the estimation of the population-based Youden index and its associated cut-point. In Section 3, we introduce the covariate-adjusted Youden index and its associated cut-point, and propose a model-free estimation framework based on the large margin classification for estimating the covariate-adjusted Youden index and its associated cut-point. In Section 4, numerical experiments are conducted to demonstrate the advantage of the proposed method. In Section 5, we apply the proposed method to the Pima Indians diabetes dataset. Section 6 contains some discussion, and the Appendix is devoted to technical proofs.

\section{The Youden index and optimal cut-point}

\begin{figure}[!h]
\caption{Receiver Operating Characteristic (ROC) curve with the Youden index $(J)$ displayed.}
\begin{center}
\includegraphics[width=0.75\textwidth]{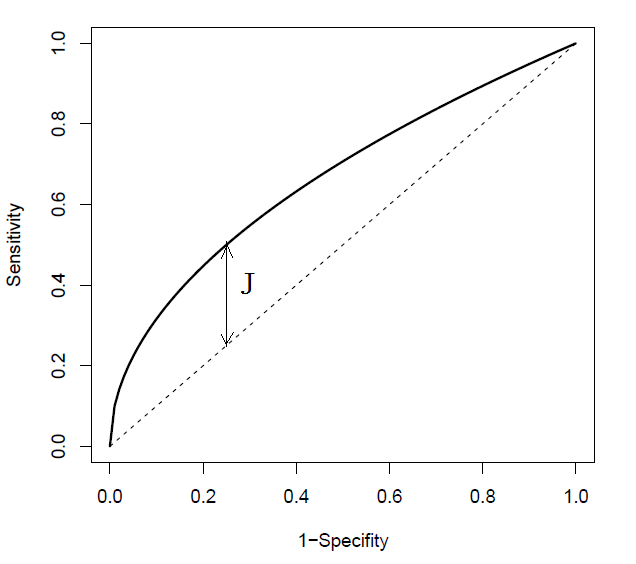}
\label{fig:Youden}
\end{center}
\end{figure}

Suppose that every observation consists of a continuously supported diagnostic measurement $X$, and a binary variable $Y \in \{-1,1\}$, where $Y = 1$ denotes a diseased subject and $Y=-1$ otherwise. A cut-point $c$ is introduced so that a diseased status is predicted if $X \geq c$ and non-diseased otherwise. The ROC curve is constructed to display the sensitivity, $\mbox{sen}(c)=Pr(X \geq c|Y=1)$ and the specificity, $\mbox{spe}(c)=Pr(X < c|Y=-1)$. To summarize the information in the ROC curve, the Youden index is defined as
$$
J= \max_c~\{\mbox{sen}(c) + \mbox{spe}(c)- 1\}.
$$
The Youden index ranges from 0 to 1, where $J=1$ represents a complete separation, and $J=0$ represents a complete overlap. The associated cut-point $c^*$ is the point that yields $J$,
$$
c^*=\argmax_c~\{\mbox{sen}(c) + \mbox{spe}(c)- 1\}.
$$
Furthermore, Figure \ref{fig:Youden} depicts the Youden index on a ROC curve \cite{Schisterman2005}.
\begin{center}
\begin{tabular}[t]{c}
      \hline
      \hline
      Figure \ref{fig:Youden} about here.\\
      \hline
      \hline
\end{tabular}
\end{center}

Since $\mbox{sen}(c)=Pr(X>c|Y=1)$ and $\mbox{spe}(c)=Pr(X \leq c|Y=-1)$, direct derivation yields that $c^*$ is a solution of
\begin{equation}
\max_c~ E\Big(w(Y)\big(1 + Y \sign(X-c)\big)\Big),
\label{eqn:optimalcut}
\end{equation}
where $w(1)=1/\pi$ and $w(-1)=1/(1-\pi)$ with $\pi=Pr(Y=1)$, and $\sign(u) = 1$ if $u \geq 0$ and $-1$ otherwise for convenience.

{\pro
\label{lem:bayes}
Assume that $p(x) = Pr(Y=1|X=x)$ is increasing in $x$, then the solution of (\ref{eqn:optimalcut}) satisfies $p(c^*) = \pi$. 
}

Furthermore, it can be showed that $c^*$ also satisfies that $f_1(c^*)=f_{-1}(c^*)$ \cite{Schisterman2005}, where $f_1$ and $f_{-1}$ are probability density functions of $x$ conditional on $Y=1$ and $Y=-1$, respectively. Note that it is important to understand how the study is designed before interpreting Theorem \ref{lem:bayes}. Different designs lead to different meanings of $\pi$, $p(x)$, $f_1$ and $f_{-1}$. Two popular designs in medical research are case-control study and cohort study. In a case-control study, the diseased status is known when sampled, and then $\pi$ is known and equal to the proportion of diseased subjects among the sampled subjects. Also, $p(x)=Pr(Y=1|X=x; {\rm{sampled}})$, $f_1$ and $f_{-1}$ are the distributions of $X$ among the diseased and non-diseased subjects, respectively. In a cohort study, $(X_i, Y_i)$, $i=1, \cdots, n$, are i.i.d., and then $\pi$ is the prevalence of the disease and can be estimated by the proportion of diseased subjects among the sampled subjects.

To estimate the Youden index and its associated cut-point, various modeling strategies have been proposed. The parametric methods \cite{Zou2000,Faraggi2002,Schisterman2005} impose distributional assumptions on $f_1$ and $f_{-1}$, and estimate $c^*$ as the solution of $f_1(c^*)=f_{-1}(c^*)$. The nonparametric methods \cite{Hsieh1996,Zou1998,Knight2000} relax the distributional assumption and estimate $f_1$ and $f_{-1}$ in a nonparametric fashion, and then estimate $c^*$ as the solution of $\hat f_1(c^*)=\hat f_{-1}(c^*)$.

By Proposition \ref*{lem:bayes}, rather than focusing on the conditional density estimation, the expectation in (\ref{eqn:optimalcut}) could be approximated by its empirical version. Then given the training sample $(x_i,y_i)_{i=1}^n$, the estimated $\hat{c}$ is defined as a solution of
\begin{eqnarray}
& & \max_c~ \frac{1}{n} \sum_{i=1}^n \hat w(y_i)(1 + y_i \sign(x_i-c)) \nonumber \\
&=& \max_c~ \frac{1}{|{\cal S}_1|}\sum\limits_{i \in {\cal S}_1} (1+\sign(x_i-c)) + \frac{1}{|{\cal S}_{-1}|}\sum\limits_{i \in {\cal S}_{-1}} (1-\sign(x_i-c)),
\label{eqn:optimalcut_est}
\end{eqnarray}
where $\hat w(1)=1/\hat \pi=n/|{\cal S}_1|$, $\hat w(-1)=n/|{\cal S}_{-1}|$, ${\cal S}_1 = \{i:y_i = 1\}$, ${\cal S}_{-1} = \{i: y_i = -1\}$, and $|\cdot|$ denotes the cardinality of a set. It is clear that the solution to (\ref{eqn:optimalcut_est}) is equivalent to estimating $f_1$ and $f_{-1}$ by their empirical distributions. The optimization in (\ref{eqn:optimalcut_est}) can be solved by an exhaustive search over all possible values of $c$, by noting that the objective function does not change when $c$ varies between two adjacent values of $x_i$'s. Another desirable property of the formulation (\ref{eqn:optimalcut_est}) is that it can be naturally extended to covariate-adjusted formulation, where $c$ is allowed to vary according to subject's profile.

\section{Covariate-adjusted formulation}

In many situations, the accuracy of diagnostic tests could be largely influenced by various factors such as the demographic characteristics of subjects \cite{Pepe2000,Faraggi2003} or the design characteristics of diagnostic tests \cite{Pepe1997}. For instance, Pepe \cite{Pepe1998} investigated an audiology study and found that the test accuracy is associated with auditory stimulus levels for patients.

To incorporate the effect of covariates, Faraggi \cite{Faraggi2003}, Simth and Thompson\cite{Smith1996}, and Guttman et al. \cite{Guttman1988} employed linear regression models. Pepe \cite{Pepe1998} further formulated a general regression framework to evaluate the effect of covariates. To relax the restrictive model assumptions, Pepe \cite{Pepe2000}, Cai and Pepe \cite{Cai2002} and Cai \cite{Cai2004} proposed a semi-parametric generalized linear model (GLM) for covariate-adjusted ROC curve without predicting the sensitivity and specificity at a given threshold. Yao et al. \cite{Yao2010} and Zhou \cite{Zhou2011} employed a non-parametric heteroscedastic regression model to address the covariate adjustment for the ROC curve and the related indices such as the AUC and the Youden index.

In this section, we generalize the formulation of large margin classification in (\ref{eqn:optimalcut_est}) to estimate the covariate-adjusted Youden index and cut-point, and evaluate the effect of covariates.

\subsection{Covariate-adjusted cut-point}

Let $\pi_z = Pr(Y=1|Z=z)$, and $p_z(x) = P(Y=1|X=x,Z=z)$, where $z$ denotes subject's profile. For convenience of describing the main idea, here we assume that $\pi_z = \pi$ for any $z$. This holds for cohort studies where subjects are sampled randomly. But for case-control studies, we need to assume that the cases (diseased subjects) and the controls (non-diseased subjects) are sampled with covariates $Z$ being matched. If covariate $Z$ is not matched, propensity scores \cite{Rosenbaum1983} could be used to estimate $\pi_z$.

Extending the formulation in (\ref{eqn:optimalcut}), the ideal covariate-adjusted cut-point $c^*(z)$ is a solution to
\begin{equation}
\max_{c}~ E\Big(w(Y)\big(1 + Y \sign(X-c(Z))\big)\Big),
\label{eqn:optimalcut_z}
\end{equation}
where $c$ is a function of $z$, and
the expectation is taken with respect to $(X, Y, Z)$. Similarly as in (\ref{eqn:optimalcut}), we can show that $c^*(z)$ must satisfy
\begin{equation}
p_z(c^*(z))=\pi,
\label{eqn:optimalcut_z_sol}
\end{equation}
where $p_z(x)$ is assumed to be a continuous and strictly increasing function of $x$ for any value of $z$.

To estimate the covariate-adjusted $c^*(z)$, note that the empirical version of (\ref{eqn:optimalcut_z})
\begin{equation}
\min_{c}~ \frac{1}{n} \sum_{i=1}^n \Big(\hat{w}(y_i)(1-y_i \sign(x_i-c(z_i)))\Big),
\label{eqn:optimalcut_z_est}
\end{equation}
involves the 0-1 loss $L_{01} (u)=\frac{1}{2}(1-\sign(u))$ and needs to be optimized with respect to functional $c(z)$. It can no longer be solved by the exhaustive grid search or any other efficient optimization techniques.
In this paper, we propose a novel surrogate loss, $\psi_{\delta}$-loss, which extends the $\psi$-loss \cite{Shen2003,Liu2006} by introducing a parameter $\delta$ that controls the difference between the surrogate loss and the 0-1 loss. More importantly, Proposition \ref{lem:consistency} shows that the $\psi_{\delta}$-loss is asymptotically Fisher consistent in estimating $c^*(z)$ when $\delta$ approaches 0.

{\pro
\label{lem:consistency}
For any given $z$, if the conditional density $f_z(x)$ is continuous and $p_z(x)$ is strictly increasing in $x$, then $E \Big ( w(Y) L_{\delta}( Y(X-c))|Z=z \Big ) \rightarrow E \Big (w(Y) L_{01}( Y(X-c) )|Z=z \Big )$ as $\delta \rightarrow 0 $ uniformly over a compact set ${\cal D}_z$ containing $c^*(z)$ and
$$
\argmin_{c} E \Big ( w(Y) L_{\delta}( Y(X-c(z)) ) |Z=z\Big ) \longrightarrow c^*(z).
$$}

With the $\psi_{\delta}$-loss, the proposed estimation formulation for the covariate-adjusted cut-point $\hat c(z)$ is a solution of
\begin{equation}
\min_{c \in {\cal F}} \frac{1}{n} \sum_{i=1}^n \hat{w}(y_i) L_{\delta}( y_i(x_i-c(z_i)) ) + \lambda {\cal J}(c),
\label{eqn:optimalcut_z_cost}
\end{equation}
where $\lambda$ is a tuning parameter, ${\cal J}(c)$ is a regularization term on the complexity of $c(z)$, and $\cal F$ is set as a reproducing kernel Hilbert space ${\cal H}_K$ (RKHS; \cite{Wahba1990}).
The final estimation formulation then becomes
\begin{equation}
\min_{c \in {\cal H}_K}~\frac{1}{n} \sum_{i=1}^n \hat{w}(y_i) L_{\delta}( y_i(x_i-c(z_i)) )  + \frac{\lambda}{2} \|c\|^2_{{\cal H}_K},
\label{eqn:rkhs}
\end{equation}
where ${\cal H}_K$ is the RKHS induced by some pre-specified kernel function $K(\cdot,\cdot)$ such as linear kernel or Gaussian kernel, and ${\cal J}(c)=\frac{1}{2}\| c
\|_{{\cal H}_K}^2$ is the associated RKHS norm of $c(z)$. It follows from the representer theorem \cite{Wahba1990} that the solution to (\ref{eqn:rkhs}) is of the form $\hat c(z) = b + \sum_{i=1}^n a_i K(z_i,z)$, and thus $\| c
\|_{{\cal H}_K}^2=a^T {\bf K} a$ with $a=(a_1,\cdots,a_n)^T$ and ${\bf K}=(K(z_i,z_j))_{i,j=1}^n$.

\subsection{Non-convex optimization}

Note that the objective function in (\ref{eqn:rkhs}) is non-convex, and thus we employ the difference convex algorithm (DCA; \cite{An1997}) to tackle the non-convex optimization. The key idea of the DCA is to decompose the non-convex objective function into the difference of two convex functions, and then construct a sequence of subproblems by approximating the second convex function with its affine minorization function.

In particular, the $\psi_{\delta}$-loss is decomposed as
$$
L_{\delta} (u)=\min \left( \frac{1}{\delta}(\delta-u)_+,1 \right )=\frac{1}{\delta}(\delta-u)_+ - \frac{1}{\delta}(-u)_+.
$$
Then the objective function in (\ref{eqn:rkhs}) can be decomposed as $s(w)=s_1(w)-s_2(w)$, where
\begin{eqnarray*}
s(w) &=& \frac{1}{n} \sum_{i=1}^n \hat{w}(y_i) L_{\delta}( y_i(x_i-c(z_i)) ) + \frac{\lambda}{2} \|c\|^2_{{\cal H}_K}, \\
s_1(w) & = &\frac{1}{n} \sum_{i=1}^n  \hat{w}(y_i) \Big (\frac{1}{\delta}\left(\delta-y_i(x_i-c(z_i))\right)_+\Big ) + \frac{\lambda}{2}\| c \|_{{\cal H}_K}^2, \\
s_2(w) & = &\frac{1}{n} \sum_{i=1}^n  \hat{w}(y_i) \Big (\frac{1}{\delta}\left(-y_i(x_i-c(z_i))\right)_+\Big ),
\end{eqnarray*}
and $w$ is the coefficient vector for the RKHS representation of $c(z)$.

Next, the DCA constructs a sequence of decreasing upper envelop of $s(w)$ by approximating $s_2(w)$ with its affine minorization function, $s_2(w^{(k)})+\langle w-w^{(k)}, \nabla s_2 (w^{(k)})\rangle$, where $w^{(k)}$ is the estimated $w$ at the $k$-th iteration, and $\nabla s_2 (w^{(k)})$ is the subgradient of $s_2(w)$ at $w^{(k)}$. The updated $w^{(k+1)}$ is then obtained by solving
\begin{equation}
w^{(k+1)}= \argmin_w~s_1(w)-s_2(w^{(k)})-\langle w-w^{(k)}, \nabla s_2 (w^{(k)})\rangle.
\label{eqn:DCA}
\end{equation}
The updating scheme is iterated until convergence. Although the DCA cannot guarantee global optimum, it delivers a superior numerical performance as demonstrated in the extensive simulation study in \cite{Liu2005}.

\subsection{Covariate-adjusted Youden index}

For any given $z$, with the covariate-adjusted cut-point $c(z)$, the covariate-adjusted Youden index $J(z)$ is expressed as
\begin{equation}
J(z) = Pr(X \geq c(z)| Y = 1, Z=z) - Pr(X \geq c(z)| Y = -1, Z=z).
\label{eqn:covariate_Youden}
\end{equation}
Then estimation of $J(z)$ boils down to estimation of the conditional probabilities in (\ref{eqn:covariate_Youden}).

In literature, Faraggi \cite{Faraggi2003} and Smith and Thompson \cite{Smith1996} estimated the conditional probabilities assuming Weibull or normal distribution, respectively. Yao et al. \cite{Yao2010} and Zhou \cite{Zhou2011} proposed to estimate the conditional probability by using the heteroscedatic regression models without assuming any distributional assumption.
In this paper, we adopt the similar kernel smoothing method as in Yao et al. \cite{Yao2010} to overcome the lack of observations sharing the same $z$ for estimating $Pr(X \geq c(z)| Y = 1, Z=z)$ and $Pr(X \geq c(z)| Y = -1, Z=z)$.  In specific, the estimated $\hat J(z)$ is
\begin{equation}
\hat{J}(z) =  \frac{\sum\limits_{i \in {\cal S}_{-1}} 1_{(-\infty, \hat{c}(z)]} K_{h_{-1}} (z_i - z)}{\sum\limits_{i \in {\cal S}_{-1}} K_{h_{-1}} (z_i -z)} - \frac{\sum\limits_{i \in {\cal S}_1} 1_{(-\infty, \hat{c}(z)]} K_{h_1} (z_i - z)}{\sum\limits_{i \in {\cal S}_1} K_{h_1} (z_i -z)},
\end{equation}
where $h_1$ and $h_{-1}$ are bandwidths for the diseased and non-diseased subjects, and $K_h (\cdot) = (1/h) K(\cdot/h)$ is any pre-specified kernel density function.

\section{Simulation examples}

This section examines the proposed estimation method for estimating the covariate-adjusted Youden index and its associated cut-point using simulated examples. The numerical performance of the proposed covariate-adjusted estimation (CAE) method is compared against normal regression model (NRM) of Faraggi \cite{Faraggi2003} and heteroscedastic regression model (HRM) of Yao et al. \cite{Yao2010} and Zhou \cite{Zhou2011}.

For illustration, the kernel function used in all methods is set as the Gaussian kernel $K(z_1,z_2)=e^{-\|z_1-z_2\|^2/2 \sigma^2}$.
In simulated examples, the true $c(z)$ and $J(z)$ are known, and thus the empirical integrated squared errors
$$\frac{1}{n}\sum_{i=1}^n (\hat{c}(z_i)-c(z_i))^2 \quad \mbox{and} \quad \frac{1}{n}\sum_{i=1}^n (\hat{J}(z_i)-J(z_i))^2$$
are employed to select the tuning parameter $\lambda$ for estimating $c(z)$ and bandwidths $h_1$ and $h_{-1}$ for estimating $J(z)$, respectively. In all examples, the grid for selecting $\lambda$ is set as $\{10^{(s-31)/10}; s=1,\cdots,61\}$ and the grid for selecting $h$ is set as $\{10^{(s-31)/10}; s=1,\cdots,41\}$. For our method, $\delta=0.1$ for all simulated examples as discussed in {\it ``On minimum clinically important difference''} by Hedayat et al..

{\it Example 1.} A random sample $\{(X_i,Y_i,Z_i); i=1,\cdots, n\}$ is generated as follows. First, $Z_i$ is generated from $Unif(1,5)$, and $Y_i$ is generated from $Bern(1/2)$. Second, if $Y_i = -1$, then $X_i$ is generated from $N \Big(6+1.5 Z_i + 1.5 \sin(Z_i), 0.4 + \Phi(2 Z_i -6)\Big)$, where $\Phi(\cdot)$ denotes the c.d.f. of standard normal distribution. Otherwise, $X_i$ is generated from $N \Big(7.2+1.5 Z_i + 1.5 \sin(Z_i) + \sqrt{Z_i - 0.5}, 1.2 + \Phi(2 Z_i -6)\Big)$. Similar example was used in Yao et al. (2010) and Zhou (2011).

{\it Example 2.} A random sample $\{(X_i,Y_i,Z_i); i=1,\cdots, n\}$ is generated as follows. First, $Z_i$ is generated from $Unif(1,5)$, and $Y_i$ is generated from $Bern(1/2)$. Second, if $Y_i = -1$, then $X_i$ is generated from $Gamma \Big(6+1.5 Z_i + 1.5 \sin(Z_i), \sqrt{0.4 + \Phi(2 Z_i -6)}\Big)$; otherwise, $X_i$ is generated from $Gamma \Big(7.2+1.5 Z_i + 1.5 \sin(Z_i) + \sqrt{Z_i - 0.5},\sqrt{ 1.2 + \Phi(2 Z_i -6)}\Big)$.

{\it Example 3.} A random sample $\{(X_i,Y_i,Z_i); i=1,\cdots, n\}$ is generated as follows. First, $Z_i$ is generated from $N_3(\mu,I_3)$ with $\mu=(1,1,1)^T$, and $Y_i$ is generated from $Bern(1/2)$. Second, if $Y_i = -1$, then $X_i$ is generated from $N \Big(6+1.5 w^T Z_i^2 + 1.5 \sin(w^T Z_i), 0.4 + \Phi(2 w^TZ_i -6)\Big)$, where $Z_i^2=(Z_{i1}^2, Z_{i2}^2, Z_{i3}^2)^T$ and $w=(1,1,1)^T$; otherwise, $X_i$ is generated from $N \Big(7.2+1.5 w^T Z_i^2 + 1.5 \sin(w^T Z_i) + \sqrt{|w^T Z_i| }, 1.2 + \Phi(2 w^T Z_i -6)\Big)$.

{\it Example 4.} A random sample $\{(X_i,Y_i,Z_i); i=1,\cdots, n\}$ is generated as follows. First, $Z_i$ is generated from $N_3(\mu,I_3)$ with $\mu=(1,1,1)^T$, and $Y_i$ is generated from $Bern(1/2)$. Second, if $Y_i = -1$, then $X_i$ is generated from $Gamma \Big(6+1.5 w^T Z_i^2 + 1.5 \sin(w^T Z_i), \sqrt{0.4 + \Phi(2 w^TZ_i -6)}\Big)$, where $w=(1,1,1)^T$; otherwise, $X_i$ is generated from $Gamma \Big(7.2+1.5 w^T Z_i^2 + 1.5 \sin(w^T Z_i) + \sqrt{|w^T Z_i| }, \sqrt{1.2 + \Phi(2 w^T Z_i -6)}\Big)$.

In all examples, the sample size $n$ is set as $n=100, 250$ and $500$. All examples are replicated 50 times and the averaged performance and the corresponding standard deviations of empirical integrated squared errors for $\hat{c}(z)$ and $\hat{J}(z)$ are reported in Table \ref{tab:optimalcut_c(z)} and Table \ref{tab:optimalcut_J(z)}.

\begin{center}
\begin{tabular}{c}
\hline
\hline
Table \ref{tab:optimalcut_c(z)} and \ref{tab:optimalcut_J(z)} about here.\\
\hline
\hline
\end{tabular}
\end{center}

It is evident that our proposed method outperforms NRM and HRM for estimating the cut-point except for the cases in Examples 1 and the case with $n=100$ in Example 2. The performance of NRM largely depends on whether the data follows a normal distribution or not. HRM yields competitive performance in Examples 1 and 2 where the number of covariate $p = 1$, but its numerical performance appears to be less satisfactory in Examples 3 and 4 with $p = 3$. This is due to the reason that HRM requires a more complicated estimation framework when $p$ gets larger (Yao et. al, 2010). Furthermore, our proposed method also delivers competitive performance for estimating $J(z)$.

\section{Real application}

In this section, we applied our proposed method to analyze the Pima Indians diabetes study dataset. The Pima Indians diabetes study aims to evaluate the effectiveness of plasma glucose concentration in an OGTT for discriminating patients with diabetes, which is a classical and standard diagnostic test for diabetes \cite{WHO1985}. The dataset is publicly available at the University of California Irvine Machine Learning Repository ({\it http://archive.ics.uci.edu/ml/}). It was originally discussed in Smith et al. \cite{Smith1988}, and also analyzed by Zhou \cite{Zhou2011} using the heteroscedastic regression model for the covariate-adjusted Youden index.

The dataset consists of nine variables: number of times pregnant, plasma glucose concentration in an OGTT, diastolic blood pressure (mm Hg), triceps skin fold thickness(mm), 2-hour serum insulin (mu U/ml), body mass index, diabetes pedigree function, age (years), disease status variable (0 or 1). There are 268 subjects in the case group and 500 subjects in the control group. Incomplete observations with OGTT value 0 are removed for the analysis.

\begin{figure}[!h]
\caption{The estimated $\hat{c}(Age)$ and $\hat{J}(Age)$ in the Pima Indians diabetes study.}
\begin{center}
\includegraphics[width=0.75\textwidth]{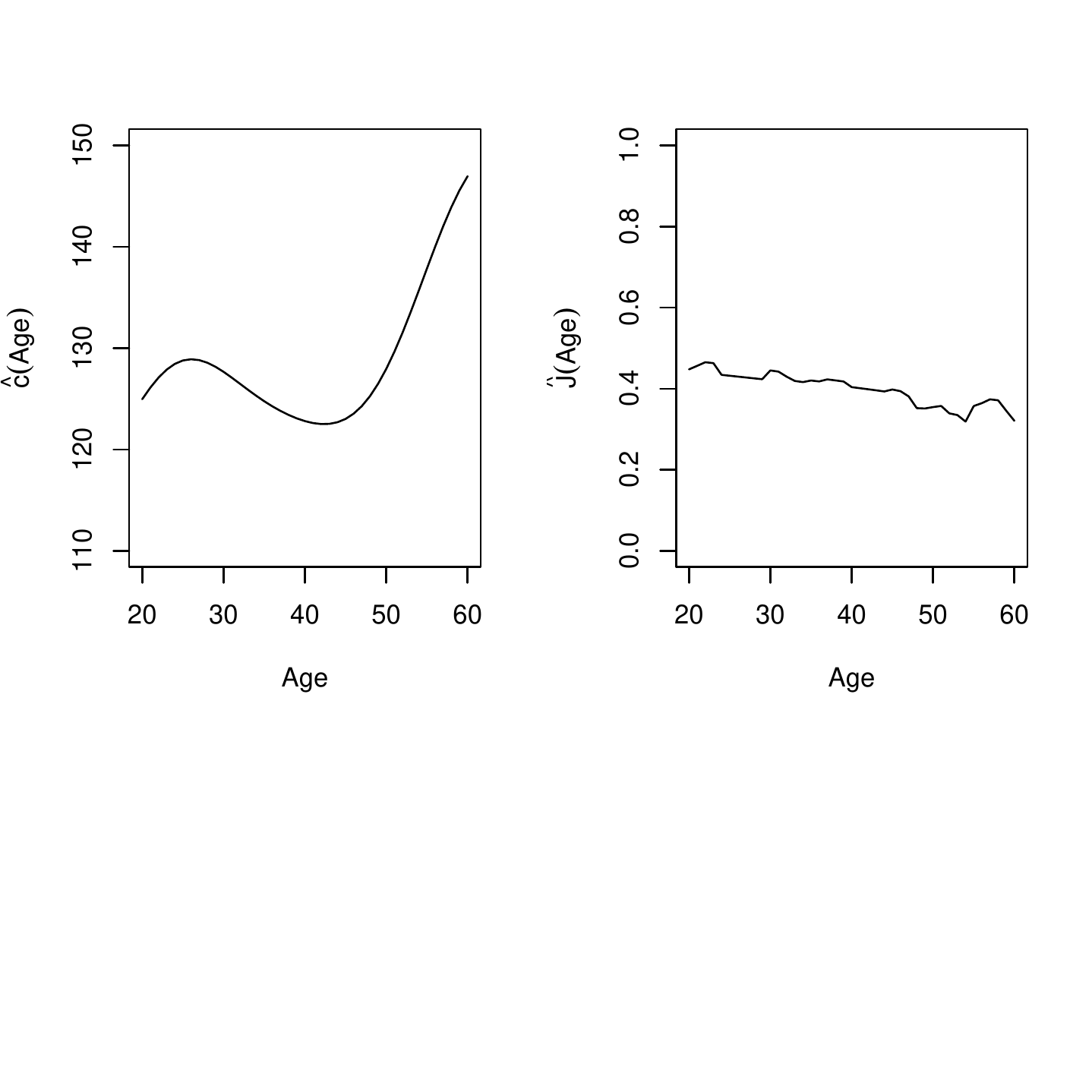}
\label{fig:real_application}
\end{center}
\end{figure}

In this section, we attempt to estimate the Youden index and the associated cut-point adjusted for the covariate age. For simplicity, we focus on the subjects who are younger than 60 years old due to the sparseness of senior subjects \cite{Zhou2011}. We also set $\delta=0.1$ and $h_{-1} = h_1 = 10$, and select the tuning parameter $\lambda$ by 5-fold cross validation targeting on maximizing (\ref{eqn:optimalcut_z_est}). Figure \ref{fig:real_application} depicts the estimated the Youden index and the associated cut-point as functions of age.

\begin{center}
\begin{tabular}{c}
\hline
\hline
Figure \ref{fig:real_application} about here.\\
\hline
\hline
\end{tabular}
\end{center}

It is clear that the value of cut-point increases with age and the value of the Youden index decreases with age. Similar conclusion has also been reached in Faraggi \cite{Faraggi2003} and Zhou \cite{Zhou2011}. Smith and Thompson \cite{Smith1996} also studied the effect of age on the ROC curve for the Cairo diabetes based on the belief that ``aging process may be associated with relative insulin deficiency or resistance among persons who do not have diabetes".


\section{Closing remarks}

This paper proposes a new framework for estimating the covariate-adjusted Youden index and its associated cut-point. As opposed to existing methods focusing on estimating conditional density functions, the proposed method targets on directly estimating the covariate-adjusted cut-point, and formulates the estimation problem in a large margin classification framework. A new surrogate loss function $\psi_\delta$ is proposed, and the resultant non-convex optimization is solved through difference convex algorithm. One key advantage of our proposed method is its estimation of the covariate-adjusted cut-point when a relatively large number of covariates are present, where multi-dimensional density estimations in existing methods are often unreliable.

\section*{Appendix}

{\bf Proof of Proposition \ref{lem:bayes}:}  Direct calculation yields that (\ref{eqn:optimalcut}) has the same solution as that of
$$
\max_c E_X \left( \frac{\sign(X-c)}{\pi(1-\pi)} (p(X)-\pi)\right).
$$
Then desired result follows immediately. That is to say,
$$
\frac{p(c^*)}{Pr(Y=1)} = \frac{1-p(c^*)}{Pr(Y=-1)},
$$
and by the Bayes' rule, $f_1(c^*) = f_{-1}(c^*)$. $\square$

\noindent {\bf Proof of Proposition 2.} For any given $z$, since $L_{\delta}(u)=L_{01} (u)+ \delta^{-1} (\delta-u)I(0 \leq u \leq \delta)$, we have
\begin{equation}
\begin{split}
E \Big ( w(Y) L_{\delta}( Y(X-c(z)) )|Z=z \Big ) & =E \Big ( w(Y) L_{01}( Y(X-c(z)) )|Z=z \Big ) \\
& + E\Big(w(Y)\frac{\delta-Y(X-c(z))}{\delta} I(0 \leq Y(X-c(z)) \leq \delta) |Z=z \Big). \label{eqn:lemma2}
\end{split}
\end{equation}
Note that $E\big(w(Y) \frac{\delta-Y(X-c(z))}{\delta} I(0 \leq Y(X-c(z)) \leq \delta) |Z=z \big)$ is decreasing in $\delta$, and approaches 0 when $\delta \rightarrow 0$. Furthermore, $E \big ( w(Y) L_{01}( Y(X-c(z)) )|Z=z \big ) - E \big ( w(Y) L_{01}( Y(X-c^*(z)) )|Z=z \big ) = \int_{c^*(z)}^{c(z)} (p_z(x)-\pi)/(\pi(1-\pi)) f_z(x) dx$, which is increasing in $c(z)$ when $c(z)>c^*(z)$. Therefore, there exist $\delta_u (z)>0$ and $c_u (z)$ such that
$$
\int_{c^*(z)}^{c_u(z)} \frac{p_z(x)-\pi}{\pi(1-\pi)} f_z(x) dx \geq E\Big( w(Y) \frac{\delta_u-Y(X-c)}{\delta_u} I(0 \leq Y(X-c) \leq \delta_u) |Z=z \Big).
$$
This implies that for any $\delta < \delta_u (z)$, $\argmin_{c} E \Big (w(Y) L_{\delta}( Y(X-c) )|Z=z \Big ) \leq c_u (z)$. Similarly, there exist $\delta_l(z)$ and $c_l (z)$ such that for any $\delta < \delta_l (z)$, $\argmin_{c} E \Big (w(Y) L_{\delta}( Y(X-c) )|Z=z \Big ) \geq c_l (z)$. Therefore, for any $\delta < \min \{\delta_l (z), \delta_u (z)\}$, $\argmin_{c} E \Big ( w(Y) L_{\delta}( Y(X-c) )|Z=z \Big )$ must lie in a compact set ${\cal D}(z)$ around $c^*(z)$.

The second term on the right hand side of (\ref{eqn:lemma2}) is bounded below by 0 and above by $\max\{ 1/\pi, 1/(1-\pi)\}P\Big(|X-c| \leq \delta |Z=z \Big)$ and is decreasing in $\delta$. Therefore, by Dini's theorem, $\max\{ 1/\pi, 1/(1-\pi)\} P\Big(|X-c| \leq \delta |Z=z \Big)$ converges to 0 uniformly over ${\cal D} (z)$ as $\delta \rightarrow 0$. It further implies that $E \Big (w(Y) L_{\delta}( Y(X-c) ) |Z=z \Big)$ converges to $E \Big ( w(Y) L_{01}( Y(X-c) ) |Z=z \Big)$ uniformly over ${\cal D} (z)$ as $\delta \rightarrow 0$. This, together with the fact that $E \Big (w(Y) L_{01}( Y(X-c) ) |Z=z \Big)$ is convex in $c$, implies that
$$
\argmin\limits_{c} E \Big ( w(Y) L_{\delta}( Y(X-c(z)) )|Z=z \Big ) \longrightarrow \argmin\limits_{c} E \Big (w(Y) L_{01}( Y(X-c(z)) )|Z=z \Big ) = c^*(z),
$$
when $\delta$ converges to zero. $\square$


{}

\newpage

\begin{table}[!ht]
      \begin{center}
      \caption{Estimated means and standard deviations (in parentheses) of empirical integrated squared errors of $c(z)$ by using our proposed method (CAE), normal regression model(NRM), and heteroscedastic regression model (HRM) based on 50 replications.}
        \begin{tabular}{cccc}
         \hline
         \hline
              &~n=100~ & ~n=250~ & ~n=500~   \\
              \hline
              & \multicolumn{2}{c}{\it Example 1} & \\
              \hline
              CAE & 0.126 (0.0887) & 0.060 (0.0401) & 0.048 (0.0398)\\
              NRM & 0.087 (0.0366) & 0.075 (0.0147) & 0.073 (0.0100) \\
              HRM & 0.069 (0.0358) & 0.051 (0.0277) & 0.035 (0.0170) \\
              \hline
              & \multicolumn{2}{c}{\it Example 2} & \\
              \hline
              CAE & 0.964 (0.9410) & 0.462 (0.3466) & 0.278 (0.1995) \\
              NRM & 1.066 (0.5920) & 0.918 (0.3896) & 0.945 (0.3284)\\
              HRM & 0.897 (0.5400) & 0.496 (0.2659) & 0.311 (0.1372)\\
              \hline
              & \multicolumn{2}{c}{\it Example 3} & \\
              \hline
              CAE & 1.715 (1.1838) & 0.796 (0.4590) & 0.441 (0.1781)\\
              NRM & 15.786 (5.4328) & 14.863 (3.7421) & 15.352 (2.6172)\\
              HRM & 3.595 (1.8378) & 2.008 (0.6182) & 1.379 (0.3460)\\
              \hline
              & \multicolumn{2}{c}{\it Example 4} & \\
              \hline
              CAE & 8.979 (3.1713)  & 5.545 (2.0988)  & 3.632 (0.9736) \\
              NRM & 25.727 (4.8747) & 20.963 (3.2498)  & 21.655 (2.5352)\\
              HRM & 10.723 (3.9732) & 6.377 (2.1690)   & 4.531 (1.0852)\\
              \hline
        \hline
        \end{tabular}
        \label{tab:optimalcut_c(z)}
       \end{center}
\end{table}

\begin{table}[!ht]
      \begin{center}
      \caption{Estimated means and standard deviations (in parentheses) of empirical integrated squared errors of $J(z)$ by using our proposed method (CAE), normal regression model(NRM), and heteroscedastic regression model (HRM) based on 50 replications.}
        \begin{tabular}{cccc}
        \hline
        &~n=100~ & ~n=250~ & ~n=500~   \\
         \hline
            & \multicolumn{2}{c}{\it EISE of $J(z)$} & \\
          \hline
            & \multicolumn{2}{c}{\it Example 1} & \\
           \hline
             CAE & 0.010 (0.0084) & 0.006 (0.0041) & 0.004 (0.0030) \\
             NRM & 0.013 (0.0097) & 0.011 (0.0056) & 0.006 (0.0035)\\
             HRM & 0.010 (0.0073) & 0.007 (0.0071) & 0.006 (0.0055) \\
                    \hline
                      & \multicolumn{2}{c}{\it Example 2} & \\
                      \hline
                      CAE & 0.011 (0.0118) & 0.005 (0.0047) & 0.003 (0.0017) \\
                      NRM & 0.016 (0.0119) & 0.013 (0.0059) & 0.014 (0.0042) \\
                      HRM & 0.017 (0.0132) & 0.010 (0.0099) & 0.007 (0.0082)\\
                      \hline
                      & \multicolumn{2}{c}{\it Example 3} & \\
                      \hline
                      CAE & 0.163 (0.0421)  & 0.141 (0.0225)  & 0.129 (0.0150)\\
                      NRM & 0.270 (0.0771) & 0.288 (0.0480)  & 0.287 (0.0308)\\
                      HRM & 0.123 (0.0343) & 0.091 (0.0250)  & 0.071 (0.0168)\\
                      \hline
                      & \multicolumn{2}{c}{\it Example 4} & \\
                      \hline
                      CAE & 0.079 (0.0330) & 0.062 (0.0158) & 0.049 (0.0089) \\
                      NRM & 0.074 (0.0398) & 0.050 (0.0227) & 0.059 (0.0127)\\
                      HRM & 0.073 (0.0209) & 0.063 (0.0139) & 0.037 (0.0070)\\
           \hline
        \hline
      \end{tabular}
       \label{tab:optimalcut_J(z)}
       \end{center}
\end{table}

\end{document}